\documentclass[sigchi]{acmart} 
\usepackage{booktabs} 
\usepackage{graphicx} 
\usepackage{amsmath}  
\usepackage{textcomp} 
\usepackage{hyperref} 
\usepackage{xurl}     

\copyrightyear{2026}
\acmYear{2026}
\setcopyright{rightsretained} 
\acmConference[CHI '26]{Proceedings of the 2026 CHI Conference on Human Factors in Computing Systems}{April 13–17, 2026}{Barcelona,  Spain}
\acmBooktitle{Co-Data Workshop at the 2026 CHI Conference on Human Factors in Computing Systems (CHI '26), April 13–17, 2026, Barcelona,  Spain}

\begin{document}

\title{Structuring Human-AI Productive Interdependence by Strategic Level of Automation Selection for Qualitative Inquiry}

\author{Feng Zhou, Jacqueline Meijer-Irons, Ambar Murillo}
\affiliation{%
  \institution{Google}
  \country{USA}
}
\renewcommand{\shortauthors}{The Authors} 

\begin{abstract}

While Large Language Models (LLMs) offer a solution to the scale-versus-depth dilemma in qualitative analysis, the paradigm of maximizing automation is fundamentally at odds with the interpretive nature of qualitative inquiry. We argue that effective Human-AI collaboration is not an automation problem, but an interdependence problem. This paper reframes the design of "co-data" systems through the lens of Interdependence Theory, proposing a formal framework to structure human-AI productive interdependence. The framework guides the selection of an appropriate Level of Automation (LoA) for different stages of the qualitative analysis process by assessing task risk and the cost of validation. We present a case study where this framework led to a deliberately interdependent workflow, fostering the calibrated trust necessary for rigorous analysis. We conclude by presenting three design principles that instantiate this framework, demonstrating how to leverage AI as a powerful partner while preserving the human researcher's irreplaceable role in the transformation process of meaning-making.

\end{abstract}



\maketitle
\section{Introduction}
Qualitative analysis is an interpretive craft. It is an iterative, reflexive process of deep reading, memoing, and meaning-making where the researcher is the primary analytical instrument \cite{Nicmanis2025}. The goal is not merely to count codes, but to construct a nuanced understanding of a human phenomenon. This craft, however, faces a crisis of scale, as the volume of qualitative data from surveys, interviews, and feedback platforms may overwhelm traditional manual methods \cite{Yan2024}.

Large Language Models (LLMs) present a tempting solution, where automation can help address the problem of scale \cite{Gao2025}. However, this automation can threaten the very nature of qualitative work. A push for full automation risks de-contextualizing data, ``flattening'' rich meaning into pre-defined categories, and sidelining the researcher's indispensable interpretive lens \cite{Feuston2021, Schroeder2025}. To avoid this, we must ask a different question: not ``How much can we automate?'' but rather ``How should we structure the interdependence between AI and the human researcher to aid researchers in leveraging the scaling capabilities of AI, while maintaining the nuance that is unique to qualitative analysis.''

We posit that effective and rigorous AI-assisted qualitative analysis is a direct function of how this interdependence between AI and researchers is managed. This paper offers a framework for structuring this productive interdependence by strategically selecting an appropriate Level of Automation (LoA) \cite{Parasuraman2000}. This choice is a calculated decision based on the risk of interpretative failure and the cost of validation for each distinct task within the qualitative workflow. We ground this framework in an industry case study, showing how we evolved a productive interdependent partnership; as well as raising questions for future research.

\section{A Framework for Structuring Productive Interdependence in Qualitative Inquiry}
We consider the nature of qualitative inquiry, following guides like Nicmanis \& Spurrier \cite{Nicmanis2025}, we acknowledge that AI cannot replace the human's role in interpretive tasks like thematic development and narrative construction. Therefore, we must necessarily move to consider how a human-AI collaboration can be effectively built for the use case of large-scale qualitative data analysis. Drawing from Interdependence Theory \cite{VanLange2015}, we argue that the researcher’s primary role is a collaboration architect who structures the productive interdependence: the rules, the tasks, and how outcomes are linked. The human researcher then engages in a transformation process, interpreting the outputs of this collaboration through their own goals, values, and contextual knowledge to create meaning. Key dimensions in this productive interdependence include complementarity of each role \cite{Hemmer2025}, capability asymmetry \cite{Hemmer2025}, causal communications \cite{Lawless2021}, and the correspondence of interests (the degree to which partners’ goals align).

To architect a productive and well-calibrated interdependence that harnesses these dimensions for cooperative transformation, we must address specific tactical questions concerning the structure of human-AI collaboration in qualitative data analysis. Recognizing that the various phases of qualitative analysis necessitate distinct levels of AI engagement, we refer to the Levels of Automation (LoA) \cite{Parasuraman2000} research. Our decision regarding the appropriate LoA for each analytical phase is guided by two critical factors:

\begin{enumerate}
    \item \textbf{Risk of Interpretive Failure:} What is the cost of an AI error to the final analysis?
    \begin{itemize}
        \item High risk tasks demand a low LoA. Data analysis tasks that are central to meaning-making (e.g., final thematic synthesis, framing the narrative) are high risk, as errors here may invalidate the entire study.
        \item Low risk tasks can accommodate a higher LoA. Repetitive, low-level tasks (e.g., open coding of a single comment) are typically lower risk and more tolerant of errors.
    \end{itemize}
    \item \textbf{Validation cost:} How difficult is it to validate the AI's work?
    \begin{itemize}
        \item High cost of validation tasks require a low LoA. For example, tasks that require deep, tacit knowledge to evaluate (e.g., ``Is this the right recommendation based on the user pain points?'')
        \item Low cost of validation tasks can accommodate a higher LoA. For example, analysis tasks that can be easily checked against source text (e.g., ``Does this quote support this insight?'')
    \end{itemize}
\end{enumerate}

\section{Case Study: A Human-AI Workflow for Qualitative Analysis}
DataSat is a large-scale internal survey program at Google designed to collect feedback from thousands of employees on their experience with over 100 internal tools. Each survey run generated about 10,000 qualitative data points—a rich but overwhelming source of user sentiment. Our challenge was to scale our data analysis, while keeping the depth of analysis, to go from data to insight in a more efficient manner. We implemented a three-phase workflow, where the human researcher and AI took complementary roles \cite{Hemmer2025} across different stages with task dependent agency and causal communications to maximize the joint performance, thus creating a productive interdependence.

\textbf{Phase 1: Structuring the Foundation.} From an Interdependence Theory perspective, this initial phase was a deliberate act of collaboration architecture. The researcher first exercised unilateral control by defining the codebook and 100 seed examples, establishing the foundational ``rules of the world'' for the AI. We then introduced asymmetrical interdependence, tasking the AI with coding the next 900 comments. In this step, the AI depended on our expert examples, while we depended on its speed. However, we identified this as high-risk because the AI's goal (pattern-matching) does not perfectly correspond with the researcher's goal (interpretive accuracy). Therefore, the final, crucial step was an act of human behavior control: we manually verified all 900 AI-coded comments. This transformation of the AI's raw output into the final ``golden dataset'' was the necessary upfront investment. It ensured the interdependence for the rest of the project was aligned with our goals, creating a trustworthy foundation for the partnership that followed.

\textbf{Phase 2: AI-Assisted Open Coding (AI as Tactical Analyst).} The AI performed high-volume open coding. This collaboration was defined by high mutual dependence with causal communication: AI depended on the human's golden dataset, while the human depended on the AI's speed. The Clue And Reasoning Prompting (CART) \cite{Sun2023} framework with Many-Shot In-Context Learning \cite{Agarwal2024} was a key rule of the collaboration with behavior control that forced the AI to produce a causal ``memo'' (Clue + Reasoning). This made the AI's ``thought process'' transparent and explainable to build shared cognition to form the necessary conditions for fostering productive interdependence \cite{Schmutz2024}. Note CART framework with many-shot in-context learning significantly improved the performance of open coding that was more effective than few-shot learning and performed comparably to fine-tuning without any model weights updating \cite{Agarwal2024}. Given the low risk of interpretative failure and low cost of verification, the AI operated at a high LoA (e.g., Level 7-10).

\textbf{Phase 3: Thematic Synthesis \& Narrative Construction (Human as Strategic Interpreter).} Here, the roles became intentionally asymmetrical based on their capabilities. The AI, operating at LoA Level 4 (suggesting one alternative), produced ``AI Insights''—drafted thematic clusters with supporting quotes. This was a tactical contribution. The human researcher retained the core strategic role, performing the crucial transformation process: evaluating AI's suggestions, performing axial and selective coding to connect themes, infusing the analysis with deep organizational context, and weaving the final narrative. The shared goal of a high-impact report created a high correspondence of interests, motivating the human researchers and other stakeholders to engage deeply with AI’s output.

\section{Discussion: Design Principles for Productive Interdependence}
\textbf{Principle 1: Mandate Bi-Directional Validation to Maintain Analytic Integrity} \\
A well-structured interdependence acknowledges that neither partner is infallible and builds in mechanisms for mutual validation.
\begin{itemize}
    \item \textit{What Worked Well:} The Human-to-AI validation loop, enforced by the CART framework, was highly effective through causal communications and inherent capability asymmetries. This form of behavior control required the AI to ``show its work,'' creating an auditable ``memo'' with scales that directly demonstrated the integrity of its process and allowed the human to easily catch subtle logical flaws.
    \item \textit{What Did Not Work Well:} The AI-to-Human validation loop was passive. When the AI struggled with an ambiguous code, it was an implicit sign of low integrity in the human-defined research instrument, but the system did not flag this explicitly.
    \item \textit{Improvement:} A better interdependence’s rule should allow the AI to actively participate in methodological refinement. It could be empowered to state: ``My confidence for classifying those qualitative responses is consistently below 50\%. Should we create new codes?'' This elevates the AI to an active partner in maintaining the quality of the analytic framework itself.
\end{itemize}

\textbf{Principle 2: Use Orienting Guidance (LoA 3-5) to to Encourage Cooperative Transformation} \\
The LoA must be chosen to create a situation with a high correspondence of interests, signaling helpful intent and inviting the human researcher to engage in a cooperative transformation process.
\begin{itemize}
    \item \textit{What Worked Well:} Framing outputs as ``AI Insights'' (Level 4 automation) was clearly cooperative. The AI proposed potential themes, but the human retained full agency to accept, reject, or re-interpret them. This preserved the crucial ``interpretive space'' for the researcher to do their work.
    \item \textit{What Did Not Work Well:} The quality of the ``AI Insights'' can be brittle. Generic insights can feel unhelpful, breaking the cooperative feeling and tempting the researcher to ignore the AI entirely.
    \item \textit{Improvement:} The system should exhibit adaptiveness by dynamically negotiating its LoA. If a user rejects several insights, the AI could initiate a repair: ``My suggestions seem too broad. Would you prefer a more focused set of sub-themes (a different Level 4 offering), or a wider array of raw concepts to work with (a shift to Level 2)?'' This moves the interaction to adaptive, mixed-initiative collaboration.
\end{itemize}

\textbf{Principle 3: Design for Asymmetry to Isolate the Transformation Process} \\
A reliable interdependence of the human-AI team requires an honest understanding of the AI's limitations. In our case study, the most significant is its inability to perform the subjective transformation process due to its lack of organizational context—its product roadmaps, historical user pain points, and priority of the organization. The LoA must be chosen to create an asymmetrical partnership that respects this boundary.
\begin{itemize}
    \item \textit{What Worked Well:} Our workflow created a clear, asymmetrical interdependence: the AI was a tactical data analyst, operating within its objective transformation process (conding text based on rule); the human was the strategic interpreter, owning the subjective transformation (creating meaning and context). This implicitly communicated the limits of the AI's ability. It could identify what was in the text, but it was the researcher’s job to understand why it mattered.
    \item \textit{What Did Not Work Well:} This separation was entirely implicit in the design. A naive user could still mistake the AI's confident tone for genuine understanding, a classic pitfall, assuming the AI understood the deep context behind user complaints about a particular feature. We relied on the human researcher to be the final gatekeeper of this context.
    \item \textit{Improvement:} The system should be designed to explicitly state the boundaries of its role. When presenting a recommendation, the AI could add a disclaimer: ``Based on the provided data, I have identified a theme of 'scattered documentation.' As an AI, I have processed this information based on the established rules. The interpretation of this finding requires a human researcher's contextual knowledge.'' This makes the limits of its ability transparent.
\end{itemize}

The three principles above are not presented as direct methods for ``designing trust'' into our system. Instead, they are practical applications of our core thesis: We treat trust (Ability, Benevolence, Integrity) \cite{Mayer1995} not as a design target, but as the emergent property of a well-structured, transparent, and methodologically sound interdependent system. In this view, trust is the outcome of a partnership whose structure makes trustworthiness observable. A researcher comes to trust their AI partner because the very design of their collaborative process consistently demonstrates the AI's helpful intent, its ability and limitations, and its transparent process. Each of the principles above, therefore, is a method for structuring the interdependence in a way that allows a specific pillar of trust to manifest. By implementing these principles, we create the conditions under which a researcher can rationally and safely develop calibrated trust in their AI partner.

This perspective, however, opens up crucial questions for future research. First, how can a system infer a researcher's trust and use it to propose a shift in the interdependent structure? Our framework proposes a static LoA for each task. A more advanced system would navigate the interdependence structure with adaptiveness. For example, if a researcher consistently accepts an AI's suggestions (a signal of high trust), the system could propose taking on more interpretive responsibility and vice versa. Second, what are the most effective system-initiated strategies for trust repair? Interdependence means failures by one partner impact the other. When the AI violates a pillar of trust (e.g., a ``hallucination'' breaches integrity), the human would naturally reduce their dependence. An effective repair might involve the AI proactively requiring more explicit human validation (increasing its own dependence) until trust is re-established, creating a safe path back to a more synergistic partnership.

\section{Conclusion}
The effective and responsible use of AI in qualitative analysis requires a paradigm shift. We move beyond the pursuit of automation and instead become ``collaboration architects'', designing for productive interdependence. This paper offered a formal framework for structuring this partnership, guiding researchers to select an appropriate LoA by assessing the interpretive risks and validation costs inherent in qualitative inquiry. By investing in foundational structures, creating asymmetrical roles, and building in mechanisms for validation, we can create Human-AI partnerships that honor the core tenets of qualitative research. The goal is not an automated analyst, but an AI-assisted interpreter—a partnership that allows us to leverage the power of LLMs for scale while ensuring the final act of meaning-making remains where it belongs: in the hands of the human.

\bibliographystyle{ACM-Reference-Format}
\bibliography{ref}

\end{document}